\documentclass[usenatbib]{mn2e}
\usepackage{natbibmnfix,graphicx}

\newcommand{\aefftsys}{\mbox{ m$^2$  K$^{-1}$}}
\newcommand{\days}{\mbox{ days}}
\newcommand{\msun}{\mbox{ M$_{\odot}$}}
\newcommand{\sfrate}{\mbox{ M$_{\odot}$ yr$^{-1}$}}
\newcommand{\sfrden}{\mbox{ M$_{\odot}$ yr$^{-1}$ Mpc$^{-3}$}}

\newcommand{\Mpc}{\mbox{ Mpc}}

\newcommand{\Mhz}{\mbox{ MHz}}
\newcommand{\hunits}{\mbox{ km s$^{-1}$ Mpc$^{-1}$}}
\newcommand{\secinv}{\mbox{ s$^{-1}$}}
\newcommand{\kel}{\mbox{ K}}
\newcommand{\mkel}{\mbox{ mK}}
\newcommand{\bq}{\begin{equation}}
\newcommand{\eq}{\end{equation}}
\newcommand{\apj}{ApJ}
\newcommand{\apjl}{ApJ}
\newcommand{\aj}{AJ}
\newcommand{\mnras}{MNRAS}
\newcommand{\aap}{A\&A}

\title[High-Redshift 21 cm Emission]{ Observing the Reionization Epoch 
Through 21 Centimeter Radiation }

\author[S.~Furlanetto, A.~Sokasian, \& L.~Hernquist]{Steven
R. Furlanetto,\thanks{Email: sfurlanetto@cfa.harvard.edu} Aaron
Sokasian,\thanks{Email: asokasian@cfa.harvard.edu} and Lars
Hernquist\thanks{Email: lhernquist@cfa.harvard.edu} \\
Harvard-Smithsonian Center for Astrophysics, 60 Garden St., Cambridge,
MA 02138}

\begin{document}

\maketitle

\begin{abstract}

We study the observability of the reionization epoch through the 21 cm
hyperfine transition of neutral hydrogen.  We use a high-resolution
cosmological simulation (including hydrodynamics) together with a fast
radiative transfer algorithm to compute the evolution of 21 cm
emission from the intergalactic medium (IGM) in several different
models of reionization.  We show that the mean brightness temperature
of the IGM drops from $\delta T_b \sim 25 \mkel$ to $\delta T_b \sim
10^{-2} \mkel$ during overlap (over a frequency interval $\Delta \nu
\sim 25 \Mhz$), while the root mean square fluctuations on small
scales drop abruptly from $\langle \delta T_b^2 \rangle^{1/2} \sim 10
\mkel$ to $\langle \delta T_b^2 \rangle^{1/2} \sim 10^{-1} \mkel$ at
the end of overlap.  We show that 21 cm observations can efficiently
discriminate models with a single early reionization epoch from models
with two distinct reionization episodes.

\end{abstract}


\section{Introduction}
\label{intro}

One of the major challenges in cosmology is to understand the first
phases of structure formation.  We do not yet know whether early
generations of sources were similar to galaxies and quasars in the
local universe or qualitatively different from them (and if so, in
what ways).  Because these sources affect the formation of all
subsequent generations of luminous objects, understanding their
characteristics is critical to any theory of structure formation.  In
many ways, the defining event of this era is ``reionization,'' when
the first sources of light ionized hydrogen in the intergalactic
medium (IGM).  The timing of this event, and its morphological
evolution, contains a wealth of information about these first sources
and about the IGM itself (e.g.,
\citealt{wyithe03,cen03,haiman03,mackey,yoshida03-semian,yoshida03-wdm}).

Observational constraints on this epoch are difficult to obtain, but
those that do exist are intriguing.  The most straightforward way to
study the ionization state of the IGM is through quasar absorption
spectra: regions with relatively large HI densities appear as
absorption troughs in the quasar spectra, creating a ``Ly$\alpha$
forest'' of absorption lines.  Spectra of $z \sim 6$ quasars selected
from the Sloan Digital Sky Survey\footnote{See http://www.sdss.org/.}
(SDSS) not only show at least one extended region of zero transmission
\citep{becker} but also imply that the ionizing background is rising
rapidly at this time \citep{fan}.  Both of these results are
consistent with a scenario in which reionization ends at $z_r \sim 6$.
Another constraint comes from measuring the fraction of the cosmic
microwave background (CMB) radiation that has been Thomson scattered
by ionized gas in the IGM.  Recent observations made by the
\emph{Wilkinson Microwave Anisotropy Probe}\footnote{See
http://map.gsfc.nasa.gov/.}  (\emph{WMAP}) imply that reionization
occurred at $z_r \ga 14$ \citep{kogut03,spergel03}.  Finally,
measurements of the temperature of the Ly$\alpha$ forest at $z \sim
2$--$4$ suggest an order unity change in the ionized fraction at $z_r
\la 10$ \citep{theuns02-reion,hui03}, although this argument is
subject to important uncertainties about HeII reionization
(e.g., \citealt{sokasian02}).  Reconciling these different
observations implies that sources must exhibit substantial (perhaps
qualitative) evolution between high redshifts and the present day
\citep{sokasian03,wyithe03,cen03,haiman03}.  A better understanding of
reionization has the potential to teach us the detailed evolutionary
history of luminous sources.

Unfortunately, learning much more from these observations promises to
be difficult.  The optical depth of the IGM to Ly$\alpha$ absorption
is $\tau_\alpha \approx 6.45 \times 10^5 x_{\rm HI} [(1+z)/10]^{3/2}$
\citep{gp}, where $x_{\rm HI}$ is the neutral fraction and where we
have assumed the currently favored cosmological parameters (see
below).  A neutral fraction $x_{\rm HI} \ga 10^{-3}$ will therefore
render the absorption trough completely black; quasar
absorption spectra can clearly probe only the very late stages of
reionization.  CMB measurements, on the other hand, depend on the
ionized gas density and thus are most sensitive to the early stages of
reionization.  However, existing observations essentially provide only
an integral constraint on the ionized gas column; distinguishing
different reionization histories with CMB observations alone promises
to be quite challenging \citep{holder03}.

It is therefore crucial to develop other probes of the high-redshift
IGM.  One possibility is to observe a transition much weaker then
Ly$\alpha$.  Perhaps the most interesting candidate is the 21 cm
hyperfine line of neutral hydrogen in the IGM
\citep{field58,field59a}.  The physics of the 21 cm transition has
been well-studied in the cosmological context (e.g.,
\citealt{scott,kumar,mmr,chen03}).  There are three methods to study
the IGM through this line.  First, we can search for emission or
absorption from the diffuse IGM and condensed minihaloes (i.e.,
collapsed objects that are unable to cool and form stars) at high
redshifts.  While the absolute signal from this gas will be swamped by
foreground emission, the signal on the sky will fluctuate because of
variations in the IGM density and neutral fraction
\citep{mmr,tozzi,iliev}.  Second, we can seek a global (all sky)
signature when the neutral emitting gas is destroyed.  Assuming that
reionization occurs rapidly, this will appear as a ``step'' in the
otherwise smooth low-frequency radio spectrum of the sky
\citep{shaver}.  Third, high-resolution spectra of powerful radio
sources at high-redshifts will reveal absorption features caused by
sheets, filaments, and minihaloes in the neutral IGM
\citep{carilli,furl-21cm}.

To date, predictions about the 21 cm signal of the diffuse IGM near
the epoch of reionization have been difficult to obtain.  This is
because such predictions require a careful treatment of ionizing
sources and radiative transfer in order to describe adequately the
transition from a neutral to an ionized medium.  Only recently has it
become possible to combine numerical radiative transfer schemes with
cosmological simulations in order to study this process in detail
\citep{gnedin00,razoumov02,ciardi03-sim,sokasian03}.  As shown by
\citet{ciardi03}, such simulations allow us, for the first time, to
realistically model the behavior of 21 cm IGM emission.  In this
paper, we analyze the simulations of \citet{sokasian03} in order to
quantify the 21 cm emission expected before, during, and after
reionization.  We focus purely on the emission signal because
absorption studies require careful treatment of both ultraviolet
radiation (above \emph{and} below the Lyman limit) and X-ray photons
\citep{chen03}, which is extremely difficult (see \citealt{carilli}
for a first attempt).  We study scenarios in which the IGM is
reionized once (at either early or late times) as well as ``double
reionization'' histories that may reconcile the SDSS and \emph{WMAP}
observations \citep{wyithe03,cen03}.  We show that 21 cm measurements
offer a promising route for distinguishing reionization scenarios that
would appear identical in CMB and quasar observations.

We briefly describe our cosmological simulations and reionization
models in \S \ref{sim}.  We then discuss the physics of 21 cm
emission, and our analysis procedure, in \S \ref{21cm}.  We present
our results in \S \ref{results} and discuss their observability and
their implications for understanding reionization scenarios in \S
\ref{discussion}.

\section{Simulations}
\label{sim}

All of our calculations are performed within the Q5 cosmological
simulation of \citet{springel03}.  This smoothed-particle
hydrodynamics (SPH) simulation uses a modified version of the GADGET
code \citep{springel01} incorporating a new conservative formulation
of SPH with the specific entropy as an independent variable
\citep{springel02}.  The simulation also includes a new description of
star formation and feedback in the interstellar medium of galaxies
\citep{springel-sf}.  Within the extended series of simulations
performed by \citet{springel03}, this model yields a converged
prediction for the cosmic star formation rate.  Using simple physical
arguments, \citet{hernquist02} have shown that the star formation rate
density evolves according to:
\bq
\dot{\rho}_\star = \dot{\rho}_\star(0) \frac{\chi^2}{1 + \alpha
(\chi-1)^3 \exp(\beta \chi^{7/4})},
\label{eq:sfr}
\eq
where
\bq
\chi(z) = \left( \frac{H(z)}{H_0} \right)^{2/3}.
\label{eq:chidefn}
\eq 
Here the constants $\alpha=0.012$, $\beta=0.041$, and
$\dot{\rho}_\star(0)=0.013 \sfrden$ are chosen by fitting the
simulation results of \citet{springel03}.\footnote{We pause here to
note an error in Figure 12 of \citet{springel03}, where the simulated
star formation rates were compared to observations.  The observational
results shown in the Figure for $0<z<2$ should be multiplied by a
factor $h$ for comparison to the simulation.  When this correction is
applied, the observed star formation rate densities \emph{decrease}
and come into closer agreement with the simulation results; thus the
simulation agrees with observations for all $z<2$ about as well as is
shown by the points at $z=0$ in the Figure.} The form of the function
is motivated by two competing processes affecting the total star
formation rate.  At high redshifts, when cooling times are short, star
formation is only limited by the growth of massive haloes.  At low
redshifts, cooling times become long because of the expansion of the
universe; thus in this regime the star formation rate density evolves
primarily because of the expansion of the universe.

The simulation assumes a $\Lambda$CDM cosmology with $\Omega_m=0.3$,
$\Omega_\Lambda=0.7$, $\Omega_b=0.04$, $H_0=100 h \hunits$ (with
$h=0.7$), and a scale-invariant primordial power spectrum with index
$n=1$ normalized to $\sigma_8=0.9$ at the present day.  These
parameters are consistent with the most recent cosmological
observations (e.g., \citealt{spergel03}).  The particular simulation
we choose has $324^3$ dark matter particles and $324^3$ SPH particles
in a box with sides of $10 h^{-1}$ comoving Mpc, yielding particle
masses of $2.12 \times 10^6 h^{-1} \msun$ and $3.26 \times 10^5 h^{-1}
\msun$ for the dark matter and gas components, respectively.

The radiative transfer is fully described in \citet{sokasian01}, so we
outline only its most significant aspects here.  We perform the
adiative transfer on a Cartesian grid with $200^3$ cells using an
adaptive ray-casting scheme \citep{abel02}.  Density fields, clumping
factors, and source characteristics are taken from outputs of the
hydrodynamic simulations.  Thus, the radiative transfer is done in a
post-processing step after the simulations have been completed, and we
neglect the dynamical feedback of the radiation on structure
formation.  We turn off periodic boundary conditions for the radiation
once the ionized volume exceeds a certain level.  After this time we
add photons that reach the edge of the box to a diffuse background.
The calculation follows the reionization of both hydrogen and helium.
Because ionizing radiation from small galaxies has a substantial
effect on reionization, we select sources aggressively (including all
groups of 16 or more dark matter particles).  At each redshift, we
correct the star formation rates in low-mass haloes to match the
converged prediction of \citet{hernquist02}.

Because the source and IGM properties are determined self-consistently
by the simulation, the major unknown parameter in simulations of
reionization is the efficiency with which each galaxy produces
ionizing photons.  This parameter includes two physical effects: the
\emph{intrinsic rate} at which newly formed stars create ionizing
photons and the fraction of such photons that are able to escape the
galaxy.  As a fiducial value, we assume an intrinsic ionizing photon
production rate of $\dot{N}_{i,{\rm fid}} \equiv 10^{53}
(\dot{M}_\star/ \sfrate) \secinv$, where $\dot{M}_\star$ is the star
formation rate, corresponding to a Salpeter initial mass function
(IMF) with solar metallicity.  We then set the actual production rate
$\dot{N}_i = f_{\rm ion} \dot{N}_{i,{\rm fid}}$.  Any value $f_{\rm
ion} \leq 1$ can be accomodated by the fiducial IMF, while $f_{\rm
ion} > 1$ will \emph{require} a change in the IMF.  For example, very
massive metal-free stars are $\sim 30$ more efficient at creating
ionizing photons than ``normal'' stars \citep{bromm-vms}.
\citet{sokasian03} found that the choice $f_{\rm ion}=0.2$ matches
data from $z \sim 6$ quasars selected through the SDSS
\citep{becker,djor01}.  However, this model implies an optical depth
to Thomson scattering significantly below the limits from recent
observations by \emph{WMAP}.  \citet{sokasian03} showed that one way
to increase the optical depth without destroying the match to the
quasar observations is to gradually increase the ionizing efficiency with
redshift in the form $f_{\rm ion}(z) = 0.20 e^{(z-6)/3}$.  Such a
model achieves overlap at $z \sim 13$.  However, because $f_{\rm ion}$
declines rapidly with cosmic time, the statistics of quasar absorbers
remain essentially unaffected (i.e., most of the extra high-redshift
photons are balanced by recombinations before $z \sim 6$).  We note
that in this model $f_{\rm ion} > 1$ for $z > 11$, indicating that the
IMF at these redshifts differs from that of local galaxies.

Another way to reconcile the quasar and CMB observations is either
partial or complete ``double reionization'' \citep{cen,wyithe}.  In
such a scenario, sources reionize the universe at $z \ga 15$.  Some
form of feedback then prevents these sources from continuing to form
and grow, rapidly reducing the production rate of ionizing photons.
For example, feedback can either prevent stars from forming in small
haloes or change the mode of star formation.  (In our notation, $f_{\rm
ion}$ essentially changes discontinuously at the moment of overlap.)
Because the recombination time at high redshifts is short compared to
the Hubble time, much of the universe recombines before ``normal''
protogalaxies can cause a second reionization event at a later time.
Note that, although both smooth and discontinuous increases in $f_{\rm
ion}$ produce large optical depths to the CMB, the qualitative
evolution of the neutral gas differs markedly between the two models.
Unfortunately, our simulation cannot resolve the haloes that are
responsible for the first reionization epoch in these scenarios.  We
therefore introduce an approximate model for double reionization by
setting the neutral fraction $x_{\rm HI} = 0.01$ throughout the
simulation volume for all $z>z_t$.  At $z=z_t$, we allow the gas to
recombine.  We set $f_{\rm ion}=0.2$ for these two models.  Note that
the simulation assumes that the ionized gas has $T=5000 \kel$.  If the
photoionized gas has a higher initial temperature, recombination may
take slightly longer (the recombination time $t_{\rm rec} \propto
T^{0.7}$).

\begin{figure}
\begin{center}
\resizebox{8cm}{!}{\includegraphics{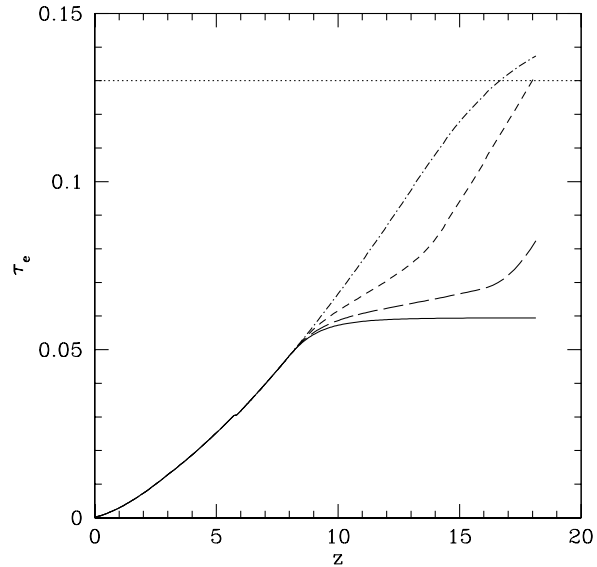}}\\%
\end{center}
\caption{ The integrated optical depth to electron scattering $\tau_e$
between the present day and redshift $z$.  The curves correspond to
our different reionization models.  The solid curve assumes a constant
$f_{\rm ion}=0.2$ for all sources throughout cosmic time.  The
dot-dashed curve assumes a variable $f_{\rm ion}(z)$ (see text).  The
short-dashed and long-dashed curves assume $f_{\rm ion}=0.2$ but set
$x_{\rm HI}=0.01$ everywhere until $z_t=14.5$ and $z_t=18.4$,
respectively. The horizontal dotted line shows the lower limit to
$\tau_e$ (at $68\%$ confidence) of \citet{kogut03}.}
\label{fig:taues}
\end{figure}

We show in Figure \ref{fig:taues} the optical depth to Thomson
scattering $\tau_e$ between the present day and redshift $z$ in these
four models.  These calculations assume that $x_{\rm HI} = x_{\rm
HeI}$ until $z=3$, when helium is instantaneously doubly ionized
throughout the universe.  The solid curve corresponds to $f_{\rm
ion}=0.2$ without an early phase of reionization.  The short- and
long-dashed curves correspond to our double reionization model, with
$z_t=14.5$ and $z_t=18.4$, respectively.  The dot-dashed curve
corresponds to the model with variable $f_{\rm ion}(z)$.  The
horizontal dotted line shows the $2\sigma$ lower limit to $\tau_e$
from \citet{kogut03}.  We note that, in the double reionization
models, we can always increase $\tau_e$ by placing the onset of the
initial reionization phase at arbitrarily high redshift.  For
reference, making the total $\tau_e=0.13$ would require the fully
ionized phase to begin by $z \sim 18.4$ or $z \sim 22$ if $z_t=14.5$ or
$z_t=18.4$, respectively.

\section{21 cm Radiation from the Intergalactic Medium}
\label{21cm}

The optical depth of the neutral IGM to redshifted hyperfine
absorption is \citep{field59a} 
\begin{eqnarray}
\tau & = & \frac{ 3 c^3 h_P A_{10} \, n_{\rm HI}}{32 \pi
k \nu_0^2 \, T_S \, H(z) } 
\label{eq:tauigm} \\
\, & \approx & 6 \times 10^{-3} (1+\delta) x_{\rm HI} \left(
\frac{T_{\rm CMB}}{T_S} \right) 
\left( \frac{\Omega_b h^2}{0.02} \right) \nonumber \\
\, & \, & \times \left[ \left(\frac{0.3}{\Omega_m} \right) \, \left(
\frac{1+z}{10} \right) \right]^{1/2} h^{-1}.
\nonumber
\end{eqnarray}
Here $h_P$ is Planck's constant, $k$ is Boltzmann's constant,
$\nu_0=1420.4 \Mhz$ is the rest-frame hyperfine transition frequency,
$A_{10} = 2.85 \times 10^{-15} \secinv$ is the spontaneous emission
coefficient for the transition, $T_S$ is the spin temperature of the
IGM (i.e., the excitation temperature of the hyperfine transition),
$T_{\rm CMB} = 2.73 (1+z) \kel$ is the CMB temperature at redshift
$z$, and $n_{\rm HI}$ is the local neutral hydrogen density.  In the
second equality, we have assumed sufficiently high redshifts such that
$H(z) \approx H_0 \Omega_m^{1/2} (1+z)^{3/2}$ (which is well-satisfied
in the era we study, $z > 6$).  We have set $\delta$ equal to the
local overdensity and $x_{\rm HI}$ equal to the neutral fraction.

In the absence of collisions and Ly$\alpha$ photon pumping, the
HI spin temperature equals the CMB temperature.  In this case
the IGM will not be visible through the 21 cm transition.  However,
either of these mechanisms can couple $T_S$ to $T_K$, the kinetic
temperature of the gas \citep{wout,field58,field59b}.  While
collisions are only effective at extremely high redshifts or in
collapsed regions, \citet{mmr} showed that even a relatively small
Ly$\alpha$ photon field is sufficient to couple the spin and kinetic
temperatures.  \citet{ciardi03} show that (in their simulations) this
requirement will be easily satisfied for the redshifts of interest.
Because our simulation has a roughly similar total emissivity to
theirs, we assume that $T_S$ and $T_K$ are closely coupled
throughout.

The visibility of the IGM then depends on its kinetic temperature.  If
$T_K \ll T_{\rm CMB}$, the IGM will appear in \emph{absorption}
against the CMB.  However, the IGM will likely be heated shortly after
the first sources appear through X-ray, photoionization, and shock
heating (e.g., \citealt{chen03}).  Once heating has occurred, the IGM
appears in \emph{emission} at a level nearly independent of $T_S$.
For simplicity, and because our simulation does not include all of the
relevant heating mechanisms (such as X-rays), we assume $T_S \gg
T_{\rm CMB}$ throughout.  At extremely high redshifts, when the first
sources are just turning on, the IGM may instead appear in absorption,
but this does not affect the main points of our paper.  In the limit
$T_S \gg T_{\rm CMB}$, the observed brightness temperature excess
relative to the CMB is \citep{mmr}
\begin{eqnarray}
\delta T_b & \approx & \frac{T_S - T_{\rm CMB}}{1+z} \, \tau 
\label{eq:dtb} \\
\, & \approx \, & 16 \, (1+\delta) x_{\rm HI} \left( \frac{T_S - T_{\rm
CMB}}{T_S} \right) \left( \frac{\Omega_b h^2}{0.02} \right) 
\nonumber \\
\, & \, & \times \left[ \left(\frac{0.3}{\Omega_m} \right) \, \left(
\frac{1+z}{10} \right) \right]^{1/2} h^{-1} \mkel.
\nonumber
\end{eqnarray}

To apply equation (\ref{eq:dtb}) to our simulation, we first choose a
bandwidth $\Delta \nu$ and beamsize $\Delta \theta$.  For reference,
these correspond to comoving distances 
\bq
L \approx 1.2 \left( \frac{\Delta \nu}{0.1 \Mhz} \right) \left(
\frac{1+z}{10} \right)^{1/2} \left( \frac{\Omega_m}{0.3}
\right)^{-1/2} h^{-1} \Mpc
\label{eq:lcom}
\eq
and
\bq
2 R \approx 1.9 \left( \frac{\Delta \theta}{1'} \right)
\left( \frac{1+z}{10} \right)^{0.2} h^{-1} \Mpc
\label{eq:rcom}
\eq 
over the relevant redshift range.  We then choose a slice of depth $L$
within the box and divide it into square pixels of width $2R$.  We
randomly select one of the axes of the box to be the line of sight and
randomly select a starting point within the box in that direction.
Next, we calculate the total mass of neutral hydrogen contained in
each pixel.  The radiative transfer code smooths each particle over
the SPH kernel and adds the appropriate mass to each cell of the
radiative transfer grid.  We multiply the total hydrogen mass in each
of these cells by the corresponding neutral fraction computed in the
simulation and then assign this mass to the appropriate pixel on the
sky.  Finally, we divide by the mean IGM mass at that redshift and apply
equation (\ref{eq:dtb}) in order to find the brightness temperature of
the pixel.  When presenting our results, we always show the average of
five such slices: because of large scale structure, a single slice is
not necessarily representative of the box as a whole.  Note that this
procedure neglects peculiar velocities of the gas particles, which
would shift some particles into or out of the frequency range we
consider.  We discuss the effects of this simplification below.

\begin{figure}
\begin{center}
\resizebox{8cm}{!}{\includegraphics{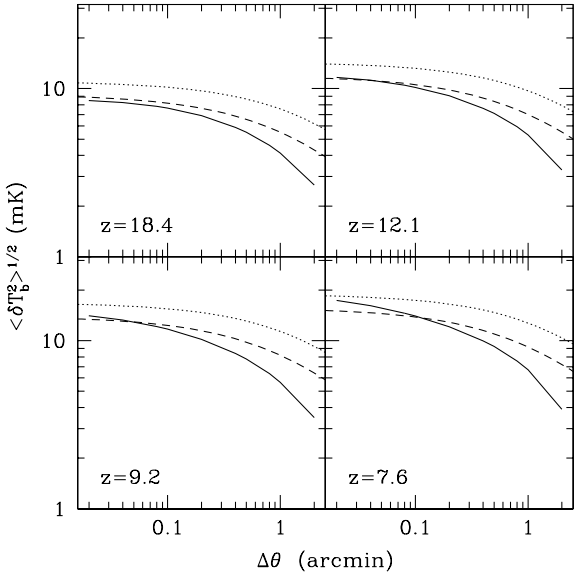}}\\%
\end{center}
\caption{ The brightness temperature fluctuations in a neutral IGM as
a function of scale at various redshifts, assuming a bandwidth $\Delta
\nu=0.1 \Mhz$.  The solid curves show the root mean square fluctuation
from the simulation assuming a fully neutral IGM.  The dashed curve
shows the estimate from equation (\ref{eq:sigmaM}) with $s(k,x)=1$.
The dotted curve shows the estimate with the \citet{kaiser87} redshift
space correction applied to all scales (see text).}
\label{fig:estimate}
\end{figure}

Although an individual IGM feature is too weak to observe, the
statistical fluctuations across the sky or in frequency space may be
observable, and they contain information about structure formation and
the ionization fraction.  Equation (\ref{eq:dtb}) shows that
fluctuations in either the density $\delta$ or neutral fraction
$x_{\rm HI}$ of the IGM lead to fluctuations in the brightness
temperature observed against the CMB.  While the full simulations are
necessary in order to compute $x_{\rm HI}$, we can estimate the level
of fluctuations due to variations in the overdensity using linear
theory.  We wish to compute the mass fluctuations inside a cylindrical
window, with the axis of the cylinder along the line of sight
\citep{tozzi}.  The result is\footnote{We note that our numerical
evaluation of equation (\ref{eq:sigmaM}) does not reproduce the
results shown in \citet{tozzi}: the predictions in their Figure 1
exceed ours by $\sim 20\%$, given identical cosmological parameters
and an identical redshift space correction.  I. Iliev has also
evaluated the integral and found results similar to ours (I. Iliev
2003, private communication).  The quantitative match with the
simulation results shown in Figure \ref{fig:estimate} gives us
confidence in our result.}  \citep{iliev}
\begin{eqnarray} 
\langle \delta T_b^2 \rangle_{\rm est} & = & \frac{8 D^2(z)}{\pi^2 R^2
L^2} \int_0^\infty dk \frac{P(k)}{k^2} \nonumber \\
& & \times \int_0^1 \frac{\sin^2
(k L x/2)}{x^2(1-x^2)} J_1^2(k R \sqrt{1-x^2}) \nonumber \\
& & \times \qquad s(k,x) dx,
\label{eq:sigmaM}
\end{eqnarray}
where $P(k)$ is the power spectrum at the present day (taken from
\citealt{eisenstein98}), $J_1(y)$ is the Bessel function of order
unity, $D(z)$ is the growth factor at redshift $z$ (normalised to
unity at the present day), and $s(k,x)$ corrects for redshift space
distortions.  In the limit in which $L \gg R$, $s(k,x) = (1 + f
x^2)^2$, where $f \approx \Omega_m^{0.6}$ \citep{kaiser87}.  We note
here that the \citet{kaiser87} correction cannot be applied outside of
this limit.  If $R \ga L$, the dominant contribution to $\langle
\delta T_b^2 \rangle_{\rm est}$ comes from modes with $k^{-1} \sim R
\ga L$, while Kaiser's derivation only applies to modes with $k^{-1}
\ll L$.  For $R \gg L$, $s(k,x)$ approaches unity.  Qualitatively,
this occurs because the small-$k$ modes sampled by such a volume are
transverse to the line of sight and hence are unaffected by peculiar
velocities.

We show how our simulation results compare to this estimate in Figure
\ref{fig:estimate}.  In each panel, the solid line shows the
simulation results for the root mean square fluctuation assuming a
fully neutral IGM, while the dashed line shows the estimate of
equation (\ref{eq:sigmaM}) with $s(k,x) = 1$ (i.e., ignoring peculiar
velocity information, as we do in our simulation analysis).  We see
that equation (\ref{eq:sigmaM}) provides a remarkably good estimate
over this entire redshift range.  Note that the simulation results
steepen slightly with cosmic time, while the shape of the estimate is
fixed [the only redshift dependence is in the growth factor $D(z)$,
which is independent of scale].  The steepening thus reflects
nonlinear structure formation.  The simulation results begin to drop
steeply for $\Delta \theta > 1'$, indicating that the simulation
cannot fully capture fluctuations on these scales.  This is not
surprising considering that the box subtends only $\theta_{\rm box}
\sim 5'$.  Comparison to equation (\ref{eq:sigmaM}) also shows that we
underestimate the fluctuations for bandwidths $\Delta \nu \ga 0.5
\Mhz$.  We focus on scales $\Delta \theta \la 1'$ and $\Delta \nu \la
0.1 \Mhz$ in most of the following.

Figure \ref{fig:estimate} also illustrates how the inclusion of
peculiar velocities would affect our results.  The dotted line shows
the estimate of equation (\ref{eq:sigmaM}) with the linear theory
redshift-space correction of \citet{kaiser87} applied blindly to all
$k$-modes.  The fluctuation signal increases by $\sim 25\%$.  Of
course, with our choice of $\Delta \nu$, this correction rigorously
only applies for $\Delta \theta \ll 1'$, and redshift space
distortions will in reality vanish as $\Delta \theta$ increases.  The
dotted lines therefore actually overestimate the effect of peculiar
velocities. We see that peculiar velocities will not affect our
qualitative results, so we chose not to complicate the analysis by
including them.  As a result, we underestimate the fluctuations by
$\la 25\%$ for angular scales small compared to the bandwidth.

\section{Results}
\label{results}

\begin{figure*}
\begin{center}
\resizebox{18cm}{!}{\includegraphics{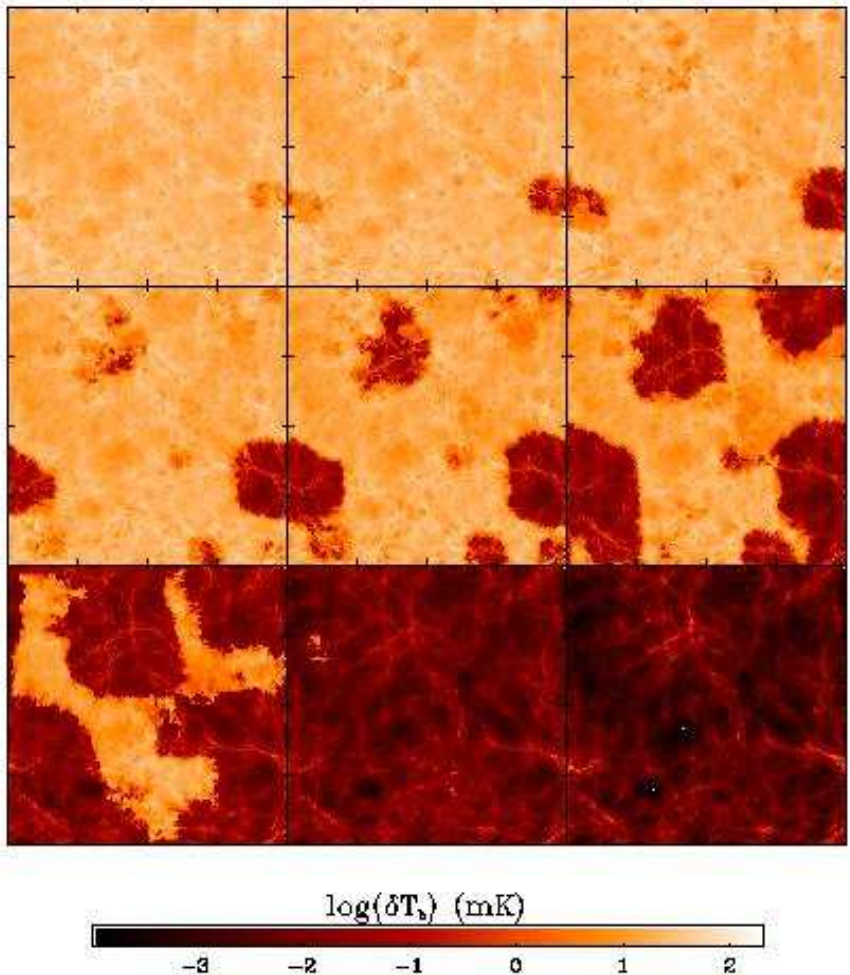}}\\%
\end{center}
\caption{ The brightness temperature from the 21 cm transition at
various redshifts in the $f_{\rm ion}=0.2$ model and with a bandwidth
$\Delta \nu=0.1 \Mhz$.  Each panel corresponds to the same slice of
the simulation (of width $10 h^{-1}$ comoving Mpc), at $z=12.1$, 11.2,
10.4, 9.8, 9.2, 8.7, 8.3, 7.9, and 7.6, from top left to bottom
right. }
\label{fig:map}
\end{figure*}

Figure \ref{fig:map} shows maps of the IGM brightness temperature
$\delta T_b$ in 21 cm emission at different redshifts.  We show the
evolution of a single slice of width $10 h^{-1}$ comoving Mpc with
$\Delta \nu = 0.1 \Mhz$ in the $f_{\rm ion}=0.2$ model.  (Note that
the angular size subtended by the box varies between redshifts.)  The
nine slices are spaced equally in cosmic time and were chosen to
include the overlap epoch.  The colourscale shows $\delta T_b$ on a
logarithmic scale.  We see the first HII regions appear and
grow, eventually filling all of space at $z \sim 8$.  Before overlap,
fluctuations arise primarily because of variations in the density of
the IGM.  After overlap, filaments stand out more clearly than their
overdensities would naively imply.  This amplification occurs because
recombination occurs more quickly in dense regions, so the residual
neutral fraction in filaments exceeds that of voids.  During overlap
the expanding HII regions produce the strongest fluctuations.

\begin{figure}
\begin{center}
\resizebox{8cm}{!}{\includegraphics{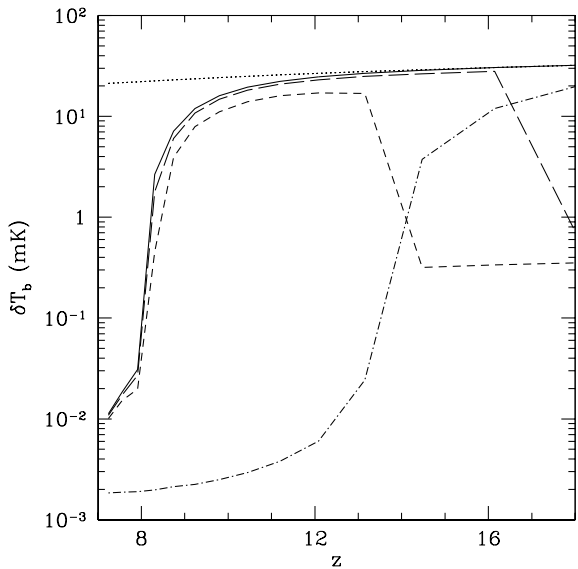}}\\%
\end{center}
\caption{ The mean brightness temperature enhancement as a function of
redshift for the different reionization models.  The dotted curve
assumes a fully neutral IGM.  The dot-dashed curve assumes a variable
$f_{\rm ion}(z)$; all other curves assume $f_{\rm ion}=0.2$.  The
short-dashed and long-dashed curves assume the IGM is $99\%$ ionized
until $z_t=14.5$ and $z_t=18.4$, respectively. }
\label{fig:meantbz}
\end{figure}

Figure \ref{fig:map} suggests that the mean $\delta T_b$ declines
dramatically during overlap.  The absolute $\delta T_b$ at a single
redshift cannot be measured because foreground emission (primarily the
CMB and Galactic synchrotron radiation) greatly exceeds the IGM
signal.  However, \citet{shaver} pointed out that this transition can
be observed as a ``step'' in frequency space, provided that overlap
occurs relatively quickly, because the foregrounds are smooth
power-laws in frequency.  We show in Figure \ref{fig:meantbz} the mean
$\delta T_b$ as a function of $z$ for each of our models.  Here the
solid line assumes $f_{\rm ion}=0.2$ and the dot-dashed line assumes a
variable ionizing efficiency $f_{\rm ion}(z)$.  The short- and long-dashed
lines assume $f_{\rm ion}=0.2$ with $z_t=14.5$ and $18.4$,
respectively.  The dotted line shows the mean signal assuming a fully
neutral IGM.  The decline at overlap ($z \sim 10$--$8$ in the $f_{\rm
ion}=0.2$ models) from $\delta T_b \sim 25 \mkel$ to near zero is
obvious.  A upward step caused by recombination after an early ionization
epoch will also be clearly visible and of similar magnitude.  Note
that the $z_t=14.5$ model never fully recombines, so its $\delta T_b$
remains somewhat smaller than the case with no early reionization
epoch.  The decline at overlap is less rapid for the variable ionizing
efficiency model, both because of the shorter recombination time during
reionization and because (by construction) $f_{\rm ion}$ decreases
with cosmic time.  A given redshift corresponds to an observer
frequency $\nu = 142 \, [10/(1+z)] \Mhz$.  In both the early and late
reionization scenarios, the fluctuations decrease by an order of
magnitude over a frequency range $\Delta \nu \sim 25 \Mhz$.

\begin{figure}
\begin{center}
\resizebox{8cm}{!}{\includegraphics{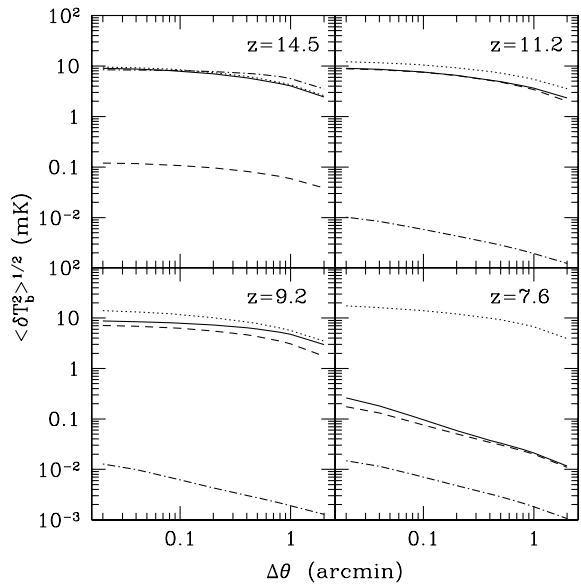}}\\%
\end{center}
\caption{ The mean brightness temperature fluctuations as a function
of scale at various redshifts, assuming a bandwidth $\Delta \nu=0.1
\Mhz$.  Each curve corresponds to a different reionization scenario: a
fully neutral IGM (dotted lines), $f_{\rm ion}=0.2$ (solid lines),
a variable $f_{\rm ion}(z)$ (dot-dashed lines), and an ionized IGM above
$z_t=14.5$ (short-dashed lines).  }
\label{fig:tscales}
\end{figure}

Unfortunately, detecting individual features in Figure \ref{fig:map}
(rather than the global transition from a neutral to an ionized IGM)
would require exceedingly long observations, even with the next
generation of radio telescopes.  Instead we may hope to measure the
statistical brightness temperature fluctuations caused by variations
in the neutral hydrogen density.  We first consider fluctuations
across the plane of the sky at a fixed frequency.  Figure
\ref{fig:tscales} shows the root mean square variation $\langle
\delta T_b^2 \rangle^{1/2}$ as a function of the beamsize at a series
of redshifts.  In each panel, the dotted line assumes a fully neutral
IGM, the solid line assumes $f_{\rm ion}=0.2$, the dot-dashed line
assumes a variable $f_{\rm ion}(z)$, and the short-dashed line assumes
$z_t=14.5$.  All assume a bandwidth $\Delta \nu = 0.1 \Mhz$.  Note
that the finite box size artificially suppresses the fluctuations on
scales $\Delta \theta > 1'$ (see Figure \ref{fig:estimate}).

Comparing the four panels, we see that the fluctuations remain
approximately constant with time until overlap is complete (at $z \sim
12$ for the variable ionizing efficiency case and $z \sim 8$ for the
other two cases).  Until this time, the fluctuation amplitude in the
models including ionizing photons is approximately the same as that of
a fully neutral medium.  At overlap, $\langle \delta T_b^2
\rangle^{1/2}$ declines sharply.  In the double reionization scenario
the fluctuations are strongly suppressed at $z > z_t$ but quickly
return to near the level of a neutral IGM.  This is simply because the
recombination time is short at high redshifts.

We therefore find that, in principle, measuring the magnitude of the
fluctuations as a function of scale at a fixed frequency is an
effective discriminant between different reionization models.
Unfortunately, fluctuations in the foregrounds present substantial
problems for actually measuring IGM fluctuations.  While Galactic
foregrounds will be smooth on these scales, extragalactic foregrounds
may not be.  \citet{dimatteo} extrapolated the known population of
low-frequency radio point sources to smaller flux levels and estimated
$\langle \delta T_b^2 \rangle_{fg}^{1/2} \ga 100$--$1000 \mkel$.
\citet{oh03} calculated the fluctuations from free-free emission by
ionized haloes at high redshifts and found $\langle \delta T_b^2
\rangle_{fg}^{1/2} \ga 100 \mkel$.  Either of these signals would be
enough to swamp the variations we predict before and during overlap,
so seeking fluctuations across the sky is in reality unlikely to
provide much information about reionization.

Fortunately, fluctuations should be easier to extract in pencil-beam
surveys of a fixed beamsize made over a large frequency range.  Here
the sampled volume is a long cylinder divided into segments by the
channel width of the radio receiver.  Nearly all known foreground
contaminants produce either free-free or synchrotron radio emission
and thus have smooth power-law spectra,\footnote{An exception would
occur if the line of sight intersects a galactic disk, which could
produce radio recombination line features.  However, these disks
should produce very small fluctuations \citep{oh03}, and in any case
recombination lines have a predictable structure.} so fluctuations in
the observed signal with frequency would be due either to IGM
structure or to variations in the beamsize with frequency.
\citet{oh03} estimate that the latter effect will (at worst) produce
fluctuations similar in magnitude to those of the IGM.  Careful
beamsize control will reduce the noise further.

\begin{figure}
\begin{center}
\resizebox{8cm}{!}{\includegraphics{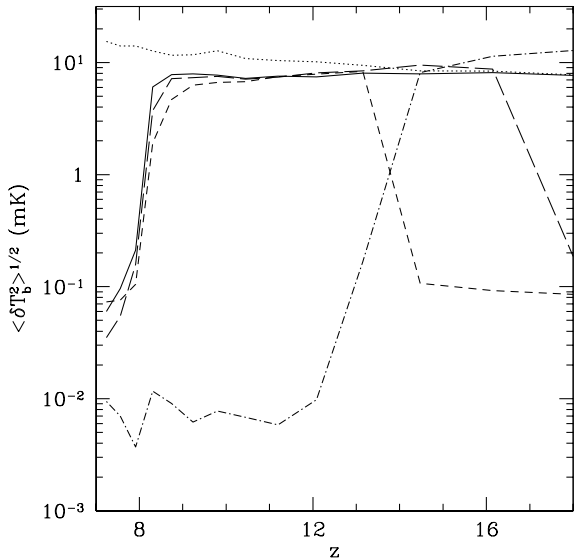}}\\%
\end{center}
\caption{ The brightness temperature fluctuations as a function of
redshift in the different reionization models, for a fixed angular
scale $\Delta \theta=0.1'$ and bandwidth $\Delta \nu=0.1 \Mhz$.  The
dotted curve assumes a fully neutral IGM.  The dot-dashed curve
assumes a variable $f_{\rm ion}(z)$; all other curves assume $f_{\rm
ion}=0.2$.  The short-dashed and long-dashed curves assume that the
IGM is $99\%$ ionized until $z_t=14.5$ and $z_t=18.4$, respectively. }
\label{fig:rmsam1}
\end{figure}

We therefore show in Figure \ref{fig:rmsam1} the root mean square
fluctuations as a function of redshift at a fixed angular scale
$\Delta \theta=0.1'$ and bandwidth $\Delta \nu=0.1 \Mhz$.  The lines
describe the same reionization models as in Figure \ref{fig:meantbz}.
In general, the behavior of $\langle \delta T_b^2 \rangle^{1/2}$ is
similar to that of $\delta T_b$: the fluctuations are large when the
medium is primarily neutral and near zero otherwise.  The transition
from a neutral to an ionized IGM (or vice versa, in the double
reionization models) is thus clearly visible as a ``step'' in the mean
level of fluctuations.

However, there are some subtle but interesting differences between the
mean signal and its fluctuations.  First, Figure \ref{fig:meantbz}
shows that in a fully neutral IGM $\delta T_b$ decreases with cosmic
time because of the expansion of the universe.  In contrast, $\langle
\delta T_b^2 \rangle^{1/2}$ increases with time because of the
continuing collapse of baryons into sheets, filaments, and bound
structures.  Second, $\langle \delta T_b^2 \rangle^{1/2}$ shows a
sharper decline than the mean signal.  The fluctuations remain large
during the early stages of overlap and fall abruptly only at the tail
end of reionization.  We find that, in all scenarios, the fluctuation
amplitude decreases by an order of magnitude over $\Delta \nu \la 15
\Mhz$.  Intuitively, this occurs because the contrast between neutral
and ionized regions dominates the fluctuations during early overlap;
until neutral regions become quite rare, the fluctuations therefore
remain large.  In the model with a variable $f_{\rm ion}(z)$, $\langle
\delta T_b^2 \rangle^{1/2}$ exceeds the fluctuations in a neutral
medium during the early stages of overlap precisely because of the
contrast between fully neutral and ionized regions.  Finally, note the
``noise'' in the estimated fluctuations after overlap, particularly in
the variable $f_{\rm ion}(z)$ model.  This indicates that, at these
times, $\langle \delta T_b^2 \rangle^{1/2}$ is more susceptible to
large scale structure variations beyond overlap (i.e., averaging five
slices is not a sufficiently representative volume past overlap).  As
in Figure \ref{fig:map}, this occurs because the shorter recombination
time in dense regions exaggerates large scale structure.  Because the
post-overlap fluctuations are unobservable even in the best cases, we
do not worry about this further.

\begin{figure}
\begin{center}
\resizebox{8cm}{!}{\includegraphics{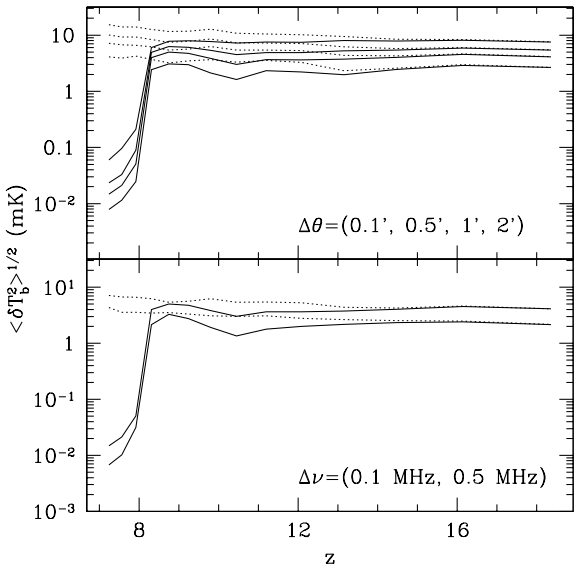}}\\%
\end{center}
\caption{ The brightness temperature fluctuations as a function of
redshift for the $f_{\rm ion}=0.2$ model.  In the top panel, we show
results for $\Delta \nu=0.1 \Mhz$ and $\Delta \theta=0.1',0.5',1.0',$
and $2.0'$, from top to bottom.  In the bottom panel, we show results
for $\Delta \theta=1.0'$ and $\Delta \nu=0.1$ and $0.5 \Mhz$, from top
to bottom.  In each panel, the dotted curves assume a fully neutral
IGM while the solid curves use the ionization fractions from the
simulation.  }
\label{fig:zvar}
\end{figure}

Figure \ref{fig:zvar} shows how the signal varies with the choice of
$\Delta \theta$ and $\Delta \nu$.  Because a larger beamsize or
bandwidth implies that each pixel samples a larger volume, the
fluctuations decrease slowly as these parameters increase.  However,
our qualitative conclusions are unaffected by the choice of scale: the
level of fluctuations remains approximately constant until the late
stages of overlap, when they decline suddenly.  In fact the decline
with scale is not quite as severe as implied by this Figure, because
the finite box size causes us to underestimate the fluctuations on
large scales by a small amount (see Figure \ref{fig:estimate}).

\section{Discussion}
\label{discussion}

We have used a high resolution hydrodynamic cosmological simulation
with the radiative transfer of ionizing photons added in a
post-processing step to examine the 21 cm emission signal from neutral
hydrogen at high redshifts.  We assume: (1) the existence of a
sufficiently strong Ly$\alpha$ background to couple the HI
spin temperature $T_S$ to the kinetic temperature $T_K$ of the IGM
\citep{mmr,ciardi03} and (2) a ``hot'' IGM satisfying $T_K \gg T_{\rm
CMB}$, with heating from, e.g., X-rays \citep{chen03}.  If these
assumptions hold, we find a mean excess brightness temperature $\delta
T_b \sim 20$--$30 \mkel$ with fluctuations $\langle \delta T_b^2
\rangle^{1/2} \sim 5$--$10 \mkel$ on small scales before reionization.
During this era, the mean signal decreases slowly with cosmic time
because of the expansion of the universe, but the fluctuations remain
approximately constant.  At overlap, both the mean signal and the
fluctuations drop rapidly to near zero.  Interestingly, while the mean
signal disappears over a redshift interval $\Delta z \ga 2$, the
fluctuations decline only in the final phase of overlap, dropping to
zero over an interval $\Delta z \la 1$.  The duration over which these
two signals fall therefore contains information about the process of
overlap.

Combining observations of high-redshift quasars \citep{becker,fan},
which suggest that reionization ended at $z_r \sim 7$, with results
from \emph{WMAP} \citep{spergel03,kogut03}, which apparently require
$z_r \ga 14$, paints an intriguing picture of reionization.
Additional constraints from the Ly$\alpha$ forest suggesting that $z_r
\la 10$ \citep{theuns02-reion,hui03} require an even more complex
reionization history (although these constraints are subject to
uncertainties about the effects of HeII reionization).
\citet{sokasian03} showed that we can reconcile the quasar and CMB
observations by smoothly increasing the efficiency of ionizing photon
production without changing the sources responsible for reionization.
In such a model, overlap occurs at $z \ga 12$, but the residual
neutral fraction remains large until $z \sim 6$ because of the short
recombination times.  Another way to reconcile these observations is
with two epochs of reionization \citep{wyithe,wyithe03,cen,cen03}.  In
these models, feedback from the sources responsible for the first
reionization event halts or modifies their subsequent formation and
growth.  The concomitant sharp drop in the efficiency of ionizing
photon production allows some fraction of the IGM to recombine and
remain mostly neutral until star-forming galaxies at $z \la 8$ ionize
it again.  Because our simulations cannot resolve the small haloes
responsible for the first era of reionization in these models, we have
approximated ``double reionization'' by simply fixing $x_{\rm
HI}=0.01$ at high redshift.  In the future, simulations of
reionization by Population III sources will allow such scenarios to be
studied in more detail (Sokasian, Yoshida, Abel, Hernquist, \&
Springel 2003, in preparation).  A different class of semi-analytic
models, which handle feedback-induced self-regulation of ionizing
sources in a perhaps more sophisticated manner, predict a long era of
partial ionization between $6 \la z \la 15$ \citep{haiman03}.

Although these reionization histories yield identical optical depths
to electron scattering and identical transmission statistics in quasar
spectra, we have shown that they can be easily distinguished through
21 cm observations.  An early overlap model causes both the mean
signal $\delta T_b$ and the fluctuations $\langle \delta T_b^2
\rangle^{1/2}$ to drop to near zero at $z \ga 13$.  In a double
reionization model, on the other hand, both the mean and fluctuations
return to levels close to that of a fully neutral medium after the gas
has recombined (see Figures \ref{fig:meantbz} and \ref{fig:rmsam1}).
Our simulations do not allow us to study self-consistent models with
long eras of partial reionization.  Nevertheless, our results suggest
a clear signature for such a reionization history.  The mean signal
$\delta T_b$ is simply proportional to the ionized fraction, so it
would decline slowly with cosmic time in a model like those of
\citet{haiman03}.  On the other hand, we have found that $\langle
\delta T_b^2 \rangle^{1/2}$ remains large until the tail end of
overlap.  We therefore suggest that these models can be identified by
a slowly decreasing mean signal accompanied by large fluctuations.
Thus, all of the reionization scenarios proposed to date can be
distinguished in a straightforward manner through 21 cm emission
measurements at $\nu \sim 100$--$200 \Mhz$, corresponding to $z \sim
14$--$6$.  This should be contrasted with predictions for advanced CMB
polarization measurements, in which the differences between these
scenarios are quite subtle (see, e.g., \citealt{holder03}).

\citet{ciardi03} have recently studied 21 cm emission with the
reionization simulations of \citet{ciardi03-sim}.  They examined early
and late reionization scenarios and, like us, found an abrupt drop in
$\langle \delta T_b^2 \rangle^{1/2}$ at the end of overlap in both
cases.  We note, however, that the fluctuation level they predict
exceeds ours (and our semi-analytic estimate) by about a factor of
two.  The most likely explanation is the differing resolution of our
simulations.  Our better mass resolution allows us to follow
small-scale density fluctuations and to identify smaller ionizing
sources, which significantly alter the reionization process
\citep{sokasian03}.  Moreover, our simulation follows the gas
hydrodynamics and more accurately describes gas clumping.  On the
other hand, \citet{ciardi03} have a box twice the width of ours, so
they are better able to study fluctuations on large scales.  In any
case, this discrepancy does not affect any of the qualitative
conclusions discussed above.

Minihaloes in the IGM will also emit 21 cm radiation \citep{iliev}.
Such minihaloes are too small to be resolved by our simulation;
however, comparison of our results to the semi-analytic model of Iliev
et al.  shows that emission from the IGM will dominate that from
minihaloes provided that our assumption of $T_S \sim T_K \gg T_{\rm CMB}$
holds.  As pointed out by \citet{oh03}, this is a simple consequence of
the fact that only a small fraction of the IGM collapses into
minihaloes.  However, the high overdensities in these objects ($\delta
\ga 100$) means that the hydrogen spin temperature in minihaloes
exceeds the CMB temperature even without coupling from a diffuse
Ly$\alpha$ radiation background \citep{iliev}.  Minihalo
emission will therefore dominate before such a background has been
generated (i.e., around the time when the very first sources appear).
The Ly$\alpha$ background will rapidly build up to a point where $T_S
\sim T_K$ \citep{ciardi03}, after which emission (or absorption) from
the diffuse IGM will dominate.  During this later era, the most
promising way to study minihaloes is through absorption spectra of
high-redshift radio sources \citep{furl-21cm}.

There are two obstacles to observing the 21 cm emission signal.  The most
basic is instrumental noise.  The root mean square noise level in a
single channel of a radio telescope is
\begin{eqnarray}
\langle \delta T_b^2 \rangle^{1/2}_{\rm noise} & = & 4.5 \mkel \left(
\frac{150 \Mhz}{\nu} \right)^2 \left( \frac{2000 \aefftsys}{A_{\rm
eff}/T_{\rm sys}} \right) \nonumber \\
& & 
\left( \frac{2'}{\Delta \theta} \right)^2
\left( \frac{ 0.5 \Mhz}{\Delta \nu_{\rm ch}} \frac{10 \days}{t}
\right)^{1/2},
\label{eq:tbnoise}
\end{eqnarray}
where $A_{\rm eff}$ is the effective area of the telescope,
$T_{\rm sys}$ is its system temperature, $\Delta \nu_{\rm ch}$ is the
channel width, $t$ is the integration time, and where we have assumed
two orthogonal polarizations.  We have used the expected instrumental
parameters for the \emph{Square Kilometer Array}.\footnote{See, e.g.,
http://www.usska.org/main.html.}  The \emph{Low Frequency
Array}\footnote{See http://www.lofar.org/index.html.} is expected to
have $A_{\rm eff}/T_{\rm sys} \sim 4.5 \times 10^2 \aefftsys$ in the
relevant frequency range.  We therefore see that, even for these
advanced instruments, detecting emission from the high-redshift IGM
requires large bandwidths and/or large beamsizes.  Although our
simulation box is not large enough to predict $\langle \delta T_b^2
\rangle^{1/2}$ on the relevant scales quantitatively, we have used a
semi-analytic estimate to show that the fluctuations from density
variations decline relatively slowly with scale (see Figure
\ref{fig:estimate}).  We have further shown that this estimate is
quantitatively accurate (to $\la 30\%$) over the entire range of
redshifts and scales we study.  Our qualitative conclusions will
therefore likely not change on the larger scales that can be
realistically probed.

The currently unknown foregrounds constitute the second obstacle to
measuring this signal.  The mean brightness temperature will be
swamped by the CMB, extragalactic sources in the beam, and Galactic
synchrotron radiation.  However, because all these sources have smooth
power-law spectra, \citet{shaver} have shown that the abrupt decline
in the signal at reionization can be observed with even a moderately
sized radio telescope.  Because we seek only the mean signal in this
case, the beamsize can be made arbitrarily large (although a small
bandwidth is still desirable for learning about the duration of
overlap).  Assuming that our volume is representative, our predictions
for the change in the mean signal $\delta T_b$ will not suffer from
errors due to the finite box size.

\citet{dimatteo} and \citet{oh03} have argued that the fluctuations on
the plane of the sky caused by unresolved extragalactic point sources
and free-free emission from ionized haloes will exceed those from 21 cm
emission by at least an order of magnitude.  Measuring $\langle \delta
T_b^2 \rangle^{1/2}$ as a function of angular scale at a single
frequency is therefore unlikely to be feasible.  However, again
because all relevant foreground sources have smooth spectra, detecting
fluctuations as a function of frequency along a given line of sight
will be possible, so long as the variation in beamsize with frequency
can be controlled (see \citealt{oh03}).  Because the fluctuations are
relatively constant for transverse scales smaller than the length
corresponding to the specified bandwidth, the optimal beamsize will
have $2R \sim L$ (see equations [\ref{eq:lcom}] and [\ref{eq:rcom}]).

Each of these techniques offers a route to learn about reionization
that is complementary to quasar absorption spectra and CMB
polarization observations.  Measurements of either the mean signal or
fluctuations about the mean will tell us about the history of neutral
gas in the IGM, and combining the two will tell us about the overlap
process.  Although the observations will be technically challenging,
the reward -- the opportunity to distinguish otherwise nearly
degenerate reionization histories -- is great.

\section*{Acknowledgments}

We would like to thank V. Springel for providing the cosmological
simulations upon which this paper is based and for useful comments on
the manuscript.  We also thank I. Iliev for helpful discussions about
equation (\ref{eq:sigmaM}) and B. Robertson for managing the computer
system on which these calculations were performed.  This work was
supported in part by NSF grant AST 00-71019.  The simulations were
performed at the Center for Parallel Astrophysical Computing at the
Harvard-Smithsonian Center for Astrophysics.


\end{document}